\documentclass[aps,prl]{revtex4}
\def\ee{\end{equation}}
\def\bea{\begin{eqnarray}}

\def\ket#1{| #1\rangle}

\begin{document}

\title{Does it Make Sense to Speak of
Self-Locating Uncertainty in the Universal Wave
Function? Remarks on Sebens and Carroll}

\author{Adrian \surname{Kent}}
\affiliation{Centre for Quantum Information and Foundations, DAMTP, Centre for
  Mathematical Sciences, University of Cambridge, Wilberforce Road,
  Cambridge, CB3 0WA, U.K.}
\affiliation{Perimeter Institute for Theoretical Physics, 31 Caroline Street North, Waterloo, ON N2L 2Y5, Canada.}
\email{A.P.A.Kent@damtp.cam.ac.uk} 

\date{August 2014} 

\begin{abstract}
Following a proposal of Vaidman \cite{vaidman1998,vaidman2008,vaidman2011},
Sebens and Carroll \cite{cs1,cs2} have argued that 
in Everettian (i.e. purely unitary) quantum theory,
observers are uncertain, before they complete
their observation, about which Everettian
branch they are on.  They argue further that this
solves the problem of making sense of probabilities within 
Everettian quantum theory, even though
the theory itself is deterministic.
We note some problems with these arguments.   
\end{abstract}
\maketitle
  
\section{Introduction}

Everett's idea \cite{everett1957relative} that pure unitary quantum
evolution holds at all times continues to 
fascinate many theoretical physicists.

Making scientific sense of Everett's idea is difficult,
as evidenced by the many and generally incompatible attempts 
(see e.g. \cite{dewittmany,mwbook,geroch1984everett,deutsch1996comment,zurek2009quantum,gell1990quantum,hartle1991quantum,vaidman2011}
and references therein)
to
show how unitary quantum theory explains the appearance
of a quasiclassical world and the 
apparent validity of the Born rule and Copenhagen
quantum theory, and evidenced also by the problems with
all of these attempts (see e.g.
\cite{bell2004speakable,kent1990against,albert1988interpreting,kentoneworld,albert,price}). 
There is still nothing close to a consensus
on the most promising way forward, even among
many-worlds enthusiasts.  This adds motivation
for developing alternative ways of  
formulating quantum theory (e.g. \cite{kentsolution}) 
that have the purported advantages
of many-worlds ideas --- realism, and Lorentz invariance ---
but describe a single real world, so avoiding both
the conceptual problems and the fantastic nature of
many-worlds ideas.   Still, for many, the appeal of many-worlds
ideas evidently persists.   

This note looks at recent suggestions by
Sebens and Carroll \cite{cs1,cs2}, who take up a proposal
originally made by Vaidman \cite{vaidman1998,vaidman2008,vaidman2011}. 
Sebens and Carroll 
 ``seek to provide an epistemic -- as
opposed to a decision-theoretic -- derivation of the Born rule which
connects quantum uncertainty to the sort of self-locating uncertainty
present in very large universes'' (\cite{cs1}, p. 3).  
Our aim here is not to give a full critique of their work, nor 
to address Vaidman's earlier papers directly,  
but simply to note some problems with 
Sebens and Carroll`s general approach. 
These seem to us likely sufficient
to doom their attempt to make sense of many-worlds probabilities
via self-locating uncertainty.  That said, we do not give a  
logical proof that no ideas along these lines can ever work.
Our discussion should, we hope, in any case, help readers
to focus on unresolved issues and form their own views 
about whether or how they could be resolved.  

We focus on the question of whether it is meaningful
to say that observers are uncertain about their self-location
in the universal wave function, and can meaningfully assign
probabilities to the events that they are located in one branch
or another.  We thus do not address Sebens and Carroll's 
purported derivation of the Born rule, except in the sense
that this derivation obviously fails if there are no relevant
probabilities to discuss \footnote{For a recent critique of
various proposed derivations of the Born 
rule, see \cite{kastner2014einselection}.}. 

As Sebens and Carroll note, one of the key difficulties for Everettians is
finding any well-defined role for the numbers that, in standard
quantum theory, are defined by the Born rule to be probabilities.
This is a problem since unitary quantum theory is deterministic,
and so on the face of it there is nothing to be uncertain about.
Yet, of course, standard quantum theory is universally seen as
successful precisely because it makes quantitative probabilistic predictions
that have been confirmed (or, more precisely, not statistically 
falsified) by all experiments to date.   If quantum theory simply
said that, when we
carry out a quantum experiment, in some unobservable sense every
possible outcome happens, 
and that we might observe any one of these possible outcomes, 
no one would ever have taken it
seriously as a scientific theory.  

Sebens and Carroll \cite{cs2} introduce their key idea 
within a standard toy
model of an ideal measurement of a spin $1/2$ particle, 
involving an apparatus $A$, an observer $O$ and the environment
$\omega$, whose pre-measurement states are $ 1/\sqrt{2} (
\ket{\uparrow} + \ket{\downarrow} ), \ket{A_0}, \ket{O_0}, 
\ket{\omega_0}$ respectively.  The measurement is modelled
by successive entangling unitaries, so that, neglecting
normalization, we have
\begin{eqnarray}
\ket{O_0} ( \ket{ \uparrow } + \ket{ \downarrow }) \ket{A_0}
\ket{\omega_0} 
& \rightarrow & 
\ket{O_0} ( \ket{ \uparrow } \ket{A_{\uparrow}} + 
\ket{ \downarrow } \ket{A_{\downarrow}} ) 
\ket{\omega_0}  \\ 
& \rightarrow & 
\ket{O_0} ( \ket{ \uparrow } \ket{A_{\uparrow}}
\ket{\omega_{\uparrow}} + 
\ket{ \downarrow } \ket{A_{\downarrow}} \ket{\omega_{\downarrow}} ) \label{a} \\
& = & 
\ket{O_0}  \ket{ \uparrow } \ket{A_{\uparrow}}
\ket{\omega_{\uparrow}} + 
\ket{O_0} 
\ket{ \downarrow } \ket{A_{\downarrow}}
\ket{\omega_{\downarrow}}  \label{b} \\ 
& \rightarrow &
\ket{O_{\uparrow}}  \ket{ \uparrow } \ket{A_{\uparrow}}
\ket{\omega_{\uparrow}} + 
\ket{O_{\downarrow}}
\ket{ \downarrow } \ket{A_{\downarrow}} \ket{\omega_{\downarrow}} \,
. 
\end{eqnarray}
Here we follow \cite{cs2} in highlighting the identity 
between (\ref{a}) and (\ref{b}), for clarity.  

Roughly speaking (in the sort of language that physicists ordinarily 
use when they are talking about quantum experiments and not worrying about 
conceptual rigour), then,
in this model, first the particle interacts with the apparatus,
which then interacts with the environment, and finally -- 
via direct inspection -- with the observer.
The resulting states $\ket{X_{\uparrow}}$ and $\ket{X_{\downarrow}}$
are taken to be orthogonal, for $X = O, A$ or $\omega$.  

Of course, to examine what role, if any, probability might
have in Everettian quantum theory one needs to be much more
careful than this in explaining how physical events might be
associated with the mathematics.
So we turn now to the words Sebens and Carroll 
attach to these equations.  
First, they state that at 
the second (and third) line, 
``the wave function has
branched -- decoherence has occurred, as indicated by the 
different environment states".   

There is more that
should be explained here.  Is this notion
of branching inherently fuzzy?   Does it require
the relevant environment states to be precisely orthogonal
or only approximately?   Can we speak of branchings for 
any decomposition of the Hilbert space into tensor products,
or is it only defined for observers and environments, and
if the latter, what precisely (or at least approximately) is an observer
and an environment?  

Still, one could read the quoted statement, taken in isolation,
as roughly consistent with other modern Everettian ideas 
(e.g. \cite{wallaceontology,wallace2012emergent}),
in which branches are not fundamental, nor precisely defined, but concepts
we impose on the universal wave function for our own convenience,
using approximate definitions of observer and environment.  
On this view, in more realistic models, neither 
the total number of branches, nor
the number of branches associated with any given component
of the wave function, is well defined.  
(Sebens and Carroll explicitly agree with this last
point: see for example footnote $10$ of \cite{cs1}.)
Measurement-like interactions are happening throughout space-time, and it's
a largely arbitrary choice which and how many of them one
chooses to represent explicitly in a decomposition of 
the wave function.   So long as 
the components identified are at least approximately 
orthogonal and remain so for at least a short while, we
are free to call them branches if we wish.

So far, then, arguably, perhaps, so good.   ``The wave function has
branched." could be read as just shorthand for something like ``Let's 
write the wave function in the form (\ref{b}), which gives a nice
(although not objectively significant) way of thinking about the
mathematics."  

But now Sebens and Carroll add ``The observer is still described
by a unique state $\ket{O_0}$, but there are two copies, one in
each branch.   Such an observer (who by construction doesn't
yet know the result of the measurement) is in a state of 
self-locating uncertainty."   
These are bold, and prima facie peculiar, assertions, which seem to need 
careful justification and explanation.  

To elaborate: 
   
There are, of course, two copies of the expression $\ket{O_0}$ 
{\it written} in equation (\ref{b}) above.   
The equivalent expression (\ref{a}) uses
just one copy.   
But, of course, so far these are just statements about ink on paper.  
To translate them into statements about one or more
observers, who are uncertain about
some relevant fact about their location on branches, 
requires some principled 
general account of how we 
start from the universal wave function and 
derive an ontology that includes
(at least) observers and branches. 
Moreover, Sebens and Carroll need 
this account to give particular answers: observers must be split
into copies before they
observe the relevant event; these copies must at that point in time be
located on definite branches; each must, however, be uncertain about 
the identity of their own branch. 

To see that this is likely to be a difficult project, one
need only list some obvious questions.   

Firstly, 
(why) can we be sure that the post-measurement/pre-observation state in this toy
model {\it does} represent two copies of an observer? 
(Why) can we not say, for example, looking at equation (\ref{a}), that 
it represents one observer?   On this latter
view, our sole observer has not interacted with the entangled 
particle-apparatus-environment state, and
has no uncertainty about this state, or their
`` branch location", or anything else.   This seems perfectly coherent,
and indeed much simpler and more natural than the picture 
Sebens and Carroll suggest.   Is it somehow logically flawed?
If not, what justifies the confident assertion that it is wrong, and
the two-copy picture is right?   

Secondly, what actually {\it is} the status of branches meant to be?   
Sebens and Carroll's self-locating uncertainty
seems to be meant to be an uncertainty about an objective
fact about reality: either I am the observer on the up
branch or the observer on the down branch, but I'm not sure which.   
But then how can these be possible facts about reality, when
the choice of branches is agreed to be not objective,
but rather an arbitrary mathematical choice which -- remember --
can be made in many generally incompatible
ways in generic models?   It seems as though, in a more detailed
model, we would need to say that there is no fixed number of 
copies of observers or branches: how exactly, then, 
are we supposed to speak of the
self-location problem of an observer, and what is it that they
are supposed to be uncertain about?  

Thirdly, suppose for the sake of the argument
that (contra Wallace \cite{wallaceontology,wallace2012emergent} and 
Sebens--Carroll \cite{cs1,cs2} and most other modern Everettians) we  
were able and willing to somehow postulate 
some natural choice of branching structure as objective.
Or suppose (which might be closer to Sebens and Carroll's intentions)
that we have postulates that allow
at least {\it some} facts about the branching structure to be 
declared objective.    
Even then, exactly how
and why would this imply objective facts about observers
splitting -- {\it before} they make their observations -- into
many copies?   Is this weird pre-observation splitting
of observers supposed to happen the instant
a measurement interaction is completed?  If so, with 
respect to which reference frame?  And are we happy to 
postulate a story about an underlying objective reality
that thus not only breaks Lorentz symmetry but also
proposes that measurement events superluminally cause
observer splittings?  If not, are we
supposed to postulate a picture in which some mysterious
observer-splitting influence propagates at light speed,
radiating outwards from the location of measurement
interactions?  Or were all observers split from the
moment of birth, before many of the measurements they
later observe took place, with each copy destined in
advance to see a particular set of measurement outcomes? 

And, in any of these cases, what about interactions that are
reversed through recoherence?   Or interactions
that produce approximate but not complete decoherence?  

Finally, but importantly, and related to the last questions,
how exactly are Sebens and Carroll's
self-uncertain observers supposed to evolve over time?   
Are their splits supposed to be instantaneous?   If not,
what happens during the splitting process?   Do sequences
of splits always simply subdivide pre-existing observers --
so that we have a tree-like branching in observer space --
or can they sometimes (for instance, when unobserved external
systems decohere and then recohere) recombine?

\section{Reduced density matrices and branching}

Sebens and Carroll present another version of their proposal
in \cite{cs1}, framed in terms of reduced density matrices.
Specifically, they suggest a principle they call {\bf ESP-QM}
(\cite{cs1}, p.15):

``Suppose that an experiment has just measured observable
$\hat{O}$ of system $S$ and registered some eigenvalue
$O_i$ on each branch of the wave function.  The probability
that agent $A$ ought to assign to the detector $D$ having
registered $O_i$ when the universal wave function is
$\Psi$, $P ( O_i | \Psi )$, only depends on the reduced density
matrix of $A$ and $D$, $\hat{\rho}_{AD}$:

\begin{equation}
P( O_i | \Psi ) = P( O_i | \hat{\rho}_{AD} ) \, ." 
\end{equation}

The wording of this version, Sebens and Carroll claim (\cite{cs1},
p.15), allows them ``to shake the unrealistic assumption that the number
of branches in which a certain outcome occurs is well-defined". 
As noted above, the assumption is indeed generally agreed to be 
unrealistic, so this would be good.    Again, though, in this
version, Sebens and Carroll attach their own narratives 
to equations without any principled justification. 
What is supposed to be meant by
the event ``the detector $D$ having registered $O_i$" to
which agent $A$ is supposed to assign a probability? 
It clearly isn't the event ``the detector $D$ registered $O_i$ in
at least one branch", which, given the understanding of branching 
Sebens and Carroll want to adopt, has probability one.  
Could it be the event ``the detector $D$ registered $O_i$ on
my branch"?  It seems it must be: what other options are
there?   But then we return to the formulation of 
\cite{cs2}, and all the associated problems noted above.  

\section{Summary}

Sebens and Carroll's attempts \cite{cs1,cs2} to
find a role for probabilities in many-worlds quantum theory
via self-locating uncertainty are puzzling, since as presented  
they seem to rely on simply plucking assertions from thin air. 
Fifty-seven years of sometimes careful work 
(e.g. \cite{dewittmany,mwbook,geroch1984everett,deutsch1996comment,
gell1990quantum,hartle1991quantum,bell2004speakable,albert1988interpreting})
on trying to make
scientific sense of Everettian quantum theory ought, surely, 
to have persuaded the theoretical physics community
that one cannot make useful progress this way.  
Wherever one thinks of the scientific status of many worlds
quantum theory, one cannot reasonably, at this point, think
it is so obvious how to translate equations into statements about
a many-worlds reality that arguments and explanations are redundant.  

Perhaps Sebens and Carroll are appealing    
to unarticulated beliefs about an underlying Everettian ontology 
-- some careful account of how branches and observers can be 
identified from the universal wave function --
that supports their assertions.   Unless and until
a clearer account is given, it is hard to know whether
a consistent ontology of the form Sebens and Carroll require
exists at all.   

The most familiar
example of an Everettian ontology that could perhaps be adapted
for Sebens and Carroll's purposes is Albert and Loewer's unloved
``many-minds interpretation'' \cite{albert1988interpreting}, which
essentially everyone, including Albert and Loewer, agrees is 
utterly unsatisfactory.  

Another possible option available to try to justify Sebens and Carroll`s
assertions is to search for a selection criterion
that either picks out one particular branching structure, or
a class of approximately equivalent branching structures.
The idea here would be that, given a theory of the initial
state and a unitary evolution law, one can algorithmically
define a branching structure using simple postulates.
Proposals for postulates of this type have recently been
made by Riedel et al. \cite{Riedel:2013uoa}.   
While intriguing, the ideas of \cite{Riedel:2013uoa} are at present tentative.
In particular, the branching rules studied in
\cite{Riedel:2013uoa} are inherently non-relativistic, and so give
no hint of how to address the questions we listed concerning
the consistency of Sebens and Carroll's ideas with special
relativity (let alone general relativity or quantum gravity).
The rules given in \cite{Riedel:2013uoa} are also defined
only for a pure initial state. 

Moreover, it is worth underlining again here that, if we {\it were}
able to find reasonably natural postulates that respected physical
symmetries and defined
an objective branching structure for the universal wave
function, it would be superfluous to postulate many 
independent real worlds.  It would be simpler and more natural
to postulate that precisely one of the branches is randomly
chosen (using the Born weight distribution) and realised in
nature.    

In summary, the Everettian literature to date
strongly suggests that it will be hard for Sebens and Carroll
to find any ontology supporting their ideas.
It seems to us that any
ontology that might possibly do so would be so baroque and ad hoc
that it would entirely undercut the case for the simplicity and
elegance of Everettian quantum theory. 
Of course, those who continue to see self-locating uncertainty 
as a potentially useful notion in the 
context of many worlds quantum theory will
take this as a challenge.   We hope that our discussion
has at least helped outline the scale of this challenge. 

\section{Acknowledgements}
I am grateful to Charles Bennett,
Sean Carroll, James Hartle, Ruth Kastner, Jess Riedel, 
Charles Sebens, Graeme Smith, John Smolin, Mark Srednicki,
Lev Vaidman, Michael Zwolak
and Wojciech Zurek for very helpful comments and discussions. 
I also thank Charles Bennett and Jess Riedel for
organising the August 2014 workshop ``Quantum Foundations of a Classical
Universe'', at which several of these discussions took place.  
This work was partially supported by a 
grants from the John Templeton Foundation and from FQXi, 
and by Perimeter Institute
for Theoretical Physics. Research at Perimeter Institute is supported
by the Government of Canada through Industry Canada and by the
Province of Ontario through the Ministry of Research and Innovation.

\section*{References}

\bibliographystyle{plain}
\bibliography{location}{}

\end{document}